\documentclass[english,prb,floats,superscriptaddress,usenames]{revtex4}
\usepackage[T1]{fontenc}
\usepackage[latin9]{inputenc}
\usepackage{amsmath}
\usepackage{amssymb}
\usepackage{graphicx}
\usepackage{esint}
\usepackage{lipsum}
\usepackage{mdframed}

\makeatletter

\newcommand{\be}{\begin{equation}} 
\newcommand{\ee}{\end{equation}}
\newcommand{\bea}{\begin{eqnarray}}   
\newcommand{\eea}{\end{eqnarray}}

\newcommand{\rr}{{\bf r}}
\newcommand{\qq}{{\bf q}}

\newcommand{\bGamma}{\boldsymbol{\Gamma}}

\newcommand{\F}{\boldsymbol{F}}

\newcommand{\Fu}{{\cal F}}
\newcommand{\OO}{{\cal O}}
\newcommand{\Ha}{{\cal H}}

\newcommand{\UU}{{\cal U}}

\usepackage{babel}

\makeatother

\usepackage{babel}
\begin{document}

\date{\today}

\title{Pressure and surface tension of an active simple liquid:\\
a comparison between kinetic, mechanical  and free-energy based  approaches.}

\author{Umberto Marini Bettolo Marconi}
\address{ Scuola di Scienze e Tecnologie, 
Universit\`a di Camerino, Via Madonna delle Carceri, 62032, Camerino, INFN Perugia, Italy}
\author{Claudio Maggi}
\address{Dipartimento di Fisica, Universit\`a di Roma Sapienza, I-00185, Rome, Italy}
\author{Simone Melchionna}

\address{CNR-ISC, Consiglio Nazionale delle Ricerche, 
Dipartimento di Fisica, Universit\`a La Sapienza, 
P.le A. Moro 2, 00185 Rome, Italy}

\date{\today}

\begin{abstract}

We discuss different definitions of pressure for a system 
of active spherical particles driven by a non-thermal coloured noise.
We show that mechanical, kinetic and free-energy based approaches lead to 
 the same result up to first order in the non-equilibrium expansion parameter.
The first prescription is  based on a generalisation of the kinetic mesoscopic virial equation and  
expresses the pressure exerted on the walls in terms of the average of the virial of the inter-particle forces. In the second { approach}, the pressure and the surface tension
are identified with the volume and area derivatives, respectively, of the partition function associated
with the known stationary non-equilibrium distribution of the model.  The third { method} 
is a mechanical approach and { is related} to 
the work necessary to deform the system. The pressure is obtained
by comparing the  expression of the work in terms of local stress and strain 
with the corresponding expression in terms of microscopic distribution.
{ This is  determined from} the force balance encoded in the Born-Green-Yvon equation. 
Such a method has the advantage of giving a formula for the local  pressure tensor { and the surface tension even} in inhomogeneous situations. By direct inspection, we show that the three procedures lead to the same values of the pressure, and give support to the idea that the partition function, obtained via the unified coloured noise  approximation, is more than a formal property of the system, but determines the  stationary non-equilibrium thermodynamics of the model.

 \end{abstract}

\maketitle

\section{Introduction}

Temperature, density and pressure are important and measurable quantities characterising the macro-state of a system at equilibrium. In particular, the pressure  is both a mechanical and a  thermodynamic property since it measures the force necessary to confine a system and 
plays a major role in the equation of state. It can be obtained in different
ways: a) mechanically, by measuring the average  normal force per unit of surface, that is the rate of momentum transferred to a surface per unit area, b) by kinetic theory  using the Clausius  virial theorem or c) by measuring the isothermal  variation of the Helmholtz free energy 
with respect to the volume available to the system. The three methods give the same results as far as  equilibrium systems are concerned, whereas in the case of non-equilibrium systems the situation is not so simple and such an equivalence cannot be proved in general since the  third method requires the existence and the knowledge of an off-equilibrium thermodynamic potential.
The issue is of particular importance, nowadays, when we consider the behavior of active fluids which are forced out of equilibrium by  injecting energy  at the individual particle level, in contrast to a driven system where the forcing occurs at the boundaries of the system \cite{bechinger2016,marchetti2013hydrodynamics,zottl2016emergent}.
In active fluids, is even delicate to define a pressure that is not a mechanical pressure and for instance
it has been demonstrated that the  pressure measured at a wall is not necessarily the same as the pressure
measured in the bulk of a system \cite{solon2014pressure}.
However, we can show that for a particular active fluid model it is possible to prove that the three methods give
equivalent answers.
The virial method \cite{henderson1986clausius} is particularly useful when one wishes to derive the pressure from a numerical simulation \cite{widom2002statistical,winkler2015virial}  or from a mesoscopic  description such as the Fokker-Planck equation \cite{falasco2015mesoscopic}. 
 
In thermodynamics, the pressure can be obtained as the negative of the derivative of the Helmholtz free energy, $\Fu$,  with respect to the volume at constant temperature and number of particles. In equilibrium statistical mechanics  this translates 
into the analogous volume derivative of the logarithm of the canonical partition function. In stationary  out-of-equilibrium active systems,
the existence of $\Fu$ cannot be proved in general. However, the Gaussian colored noise model in the 
unified colored noise approximation (UCNA) \cite{hanggi1995colored}
 is a notable exception. For this model, the  N-particle distribution function is known and has a form similar to the Gibbs
distribution of an equilibrium system.
Thus, in principle, we can formally define a partition function as the classical  trace of such a distribution and obtain from it
a series of derived quantities, such as a "thermodynamic" pressure.  It is then natural to ask: does this object has the required properties
and, more importantly, does it coincide with the mechanical and the virial pressure?

A third route to determine the pressure  considers the hydrodynamic equations and under conditions of zero fluxes
and the fact that the divergence of the pressure tensor must satisfy a mechanical balance
or hydrostatic equilibrium condition: it means that the internal stress of the fluid must balance
the resultant of the body forces. Such a mechanical equilibrium condition allows  determining the pressure tensor in terms of the microscopic distribution functions.

We prove that the three methods lead to the same expression of the pressure if we limit ourselves to
the first order in the non-equilibrium parameter $\tau$,  i.e. the persistence time of the 
"self-propulsion".  
At higher order, however, we show how this proof cannot be readily extended since the hydrostatic equation involves a matrix inversion which can be performed only
up to the first order in the perturbing parameter. On the other hand, the present result seems to suggest that studying active particles models, close to equilibrium, allows unifying the thermodynamic with the mechanical approach to pressure.

The goal of the present paper is to derive explicit expressions for the pressure and the surface tension for self-propelled \cite{speck2015dynamical,speck2015ideal,nikola2015active,solon2014pressure,solon2014pressure2}
particles using the Gaussian colored noise model in the UCNA approximation which has been recently studied by our group
\cite{maggi2015multidimensional,marconi2015towards,marconi2016molecular}.
The paper is organised as follows: in section \ref{Preamble} we report the main results obtained in previous 
papers and necessary for the comprehension of the following developments.
In section \ref{MVE}, we apply the mesoscopic virial equation (MVE) of Falasco et al. \cite{falasco2015mesoscopic} to derive an expression for the pressure exerted on the walls. The method employs the
Fokker-Planck equation for the distribution function and the asymptotic  constancy of some functions of the coordinates
to determine the virial of the force acting on the active fluid. 
Such an expression for the pressure is compared in section \ref{stretching} with the
derivative of the logarithm of the partition function associated
with the stationary   configurational distribution function. To compute the volume derivative we use
a volume scaling method of the free energy employed in equilibrium statistical mechanics 
to compute the pressure of an assembly of particles. The two results are seen to coincide up to first order in the
non-equilibrium parameter, a fact which is not at all trivial, being the first one  a mechanical property
and the second a statistical property connected to the existence of a thermodynamic potential of the system.
As a secondary benefit of the stretching method, we obtain a first definition of the surface tension of the active system.
In section \ref{bauslovett},
we proceed further and use the third approach, originally due to Baus and Lovett, to compute the pressure: it uses  the concept of mechanical work associated with a local deformation of the system to obtain the pressure and the surface tension. 
 By this approach, it is in principle possible to determine the components of the pressure tensor at every point of the non-uniform fluid by measuring the work necessary to produce a local deformation.
Even in this case, it turns out that the value of the predicted pressure is the same as those derived by the two previous  approaches and regarding the surface tension we obtain an agreement with the result of section  \ref{stretching}.
Corroborated by this agreement between the different methods we apply the mechanical method to the calculation of the 
pressure tensor in an anisotropic case.
Finally, in section \ref{conclusions} we come to the conclusions and perspectives.
In order to render the reading of
the present paper easier,  we created  several appendices where the derivations of some mathematical results used in the main text have been reported.

\section{Model equations}
\label{Preamble}
In the Gaussian colored noise model we consider the
spatial configurations $(\rr_1,\dots,\rr_N)$ of a system of overdamped particles subjected to a velocity dependent frictional force  $-\gamma {\bf v}$,  and to a white  Gaussian noise 
of intensity $\sqrt{D_t}$ and Gaussian coloured noise,  characterized by intensity $\sqrt{D_a}/\tau$  and 
relaxation time $\tau$, and enclosed in  a bounded region of volume $V$ under stationary conditions and in the absence of currents are approximately described by the following probability measure 
\be
P_N(\rr_1,\dots,\rr_N)=  \frac{1}{Z_N}\exp \Bigl(- \frac{\Ha( \rr_1,\dots,\rr_N)}{T_s} \Bigr) .
 \label{principal}
 \ee 
 The distribution \eqref{principal}
is valid under the so-called UCNA approximation and results from replacing the underlying coloured noise non Markovian dynamics 
with  Markovian dynamics with new and more complicated interactions    as discussed in previous works \cite{maggi2015multidimensional,marconi2015towards}.
The distribution depends on $T_s$, an effective temperature, resulting from the sum of a
 translational temperature $T_t=D_t \gamma$ and  a swim temperature $T_a=D_a\gamma$, where $D_t$ and $D_a$ are two  coefficients
 related to the  translational  and rotational  diffusion, respectively \cite{takatori2014swim,takatori2015towards,smallenburg2015swim}.
$\Ha$ is an effective Hamiltonian which is determined in terms of the actual potential energy $\UU_{total}$
of the system, detailed in the following,
and $Z_N$ is the normalisation and can be viewed as a generalisation
to the non-equilibrium stationary regime of the partition function. 
As shown hereafter, both $\Ha$ and $Z_N$ reduce to their equilibrium counterparts when the non-equilibrium parameter $\tau/\gamma \to 0$,
or in non-dimensional form when $\frac{\tau}{\gamma \sigma^2}  \epsilon\to 0$, where $\epsilon$ and $\sigma$ are typical energy and length  scales characterizing the inter-particle interactions.

In analogy with equilibrium statistical mechanics, we introduce a "thermodynamic" pressure, $p_t$, as: 
\be
p_t=T_s\frac{\partial \ln Z_N}{\partial V} 
\ee
and discuss its properties and relationship with other definitions of pressure such as
the virial approach where the pressure is defined as the external action necessary to confine the system to a bounded region or the kinetic theory approach where the pressure is identified with the rate of momentum per unit surface transferred from the colliding particles to the walls. 

The potential energy
 may comprise one body and pairwise potentials and has the form 
 \be
\UU_{total}(\rr_1,\dots,\rr_N) =\sum_i u(\rr_i)+ \sum_{i>j} w(\rr_i-\rr_j)  .
\label{utotal}
\ee
The effective potential energy,  $\Ha$, instead, results from the adiabatic elimination of the fast degrees of freedom and 
is related to $\UU_{total}$ by:
     \be
\Ha(\rr_1,\dots,\rr_N) = \UU_{total}(\rr_1,\dots,\rr_N)  +\frac{\tau}{2 \gamma}
    \sum_k^N \Bigr(\frac{\partial \UU_{total}(\rr_1,\dots,\rr_N)}{\partial \rr_k} \Bigl)^2 
    - T_s \ln  |\det ( \Gamma_{\alpha i,\beta k}(\rr_1,\dots,\rr_N)) 
    \ee
    where Greek indexes stand for Cartesian components and
 the non-dimensional friction  matrix $\Gamma$ is defined as
\be
\Gamma_{\alpha i,\beta k}(\rr_1,\dots,\rr_N)=\delta_{\alpha\beta}\delta_{ik}+\frac{\tau}{\gamma} \frac{\partial^2 \UU_{total}(\rr_1,\dots,\rr_N) }{\partial r_{\alpha i} \partial r_{\beta k}}.
\label{gammamatrix}
\ee

The explicit structure of the matrix $\Gamma_{\alpha 1,\beta k}$  follows from \eqref{utotal}:
\be
\Gamma_{\alpha i,\beta k} =
\Bigl( \delta_{\alpha\beta}+\frac{\tau}{\gamma} u_{\alpha\beta}(\rr_i)+
\frac{\tau}{\gamma}  \sum_j w_{\alpha\beta}(\rr_i-\rr_j) \Bigl)\delta_{ik}-\frac{\tau}{\gamma}   w_{\alpha\beta}(\rr_i-\rr_k)
(1-\delta_{ik}) .
\label{matrixinverse2} 
\ee
We remark that the diagonal elements contain $(N-1)$ pairwise terms whereas the off-diagonal matrix elements
just one term. The idea is to show that in the limit of a large number of particles it is possible to neglect the 
effect presence of the off-diagonal elements.

As shown elsewhere (see ref. \cite{marconi2015towards})  assuming that the off-diagonal elements of the
$\Gamma$ matrix are negligible
 the density distribution $\rho(\rr)$ satisfies an integrodifferential equation similar to the Born-Green-Yvon (BGY) equation
 of passive fluids.
By the method illustrated in appendix \ref{offdiagonal}  we find
\bea
&&
T_s  \sum_\beta   \frac{\partial }{\partial r_{\beta }}\Bigl( [ \delta_{\alpha\beta}-\frac{\tau}{\gamma} u_{\alpha\beta}(\rr)]  \rho(\rr)
-  \frac{\tau}{\gamma} \rho(\rr)  \int d\rr_2 w_{\alpha\beta}(\rr-\rr_2) \rho(\rr_2) g(\rr,\rr_2)\Bigr) \nonumber\\
&&= -\rho(\rr)  \frac{\partial u(\rr)}{\partial r_{\alpha }} 
  -\rho(\rr) \int d\rr_2  \rho(\rr_2) g(\rr,\rr_2)   \frac{ \partial w(\rr-\rr_2) }{ \partial r_{\alpha }}   
 \label{tssum}
\eea
where we introduced the pair correlation function $g(\rr,\rr')$.
We shall  rewrite  \eqref{tssum} in the more compact form:
\be
 f_\alpha(\rr)=T_s \sum_\beta \frac{\partial }{\partial r_{\beta}}\Bigl[ \tilde \Gamma^{-1}_{\alpha\beta}(\rr) \rho(\rr)  \Bigl] 
  +\rho(\rr)\int   d\rr'   \rho(\rr') g(\rr,\rr')   \frac{ \partial w(\rr-\rr') }{ \partial r_{\alpha}}   ,
\label{rho1bgy}
\ee
 where  $\tilde \Gamma^{-1}(\rr)$ is  a $d\times d$ matrix defined by:
\be
\tilde \Gamma^{-1}_{\alpha\beta}(\rr) =
  \Bigl[ \delta_{\alpha\beta}  -   \frac{\tau}{\gamma}    
  u_{\alpha\beta}(\rr)
    - \frac{\tau}{\gamma}   \int   d\rr'    \rho(\rr') g(\rr,\rr')  w_{\alpha\beta}(\rr-\rr')  \Bigl] .
    \label{smallGamma}
\ee
and $f_\alpha(\rr) =-\rho(\rr)  \frac{\partial u(\rr)}{\partial r_{\alpha}}  $ 
is the external force per unit volume.
Let us remark that eq.
\eqref{rho1bgy}  has the form of a BGY equation  and arises from the conservation of linear momentum, that is is equivalent to $\langle \frac{\partial {\bf J}}{\partial t}\rangle$, where
$ {\bf J}$ is the momentum density.

  In the present theory
the matrices $\Gamma$ and $\tilde \Gamma$ play a central role because they
describe how  the mobility of the particles depends on their environment   \cite{tailleur2008statistical,cates2014motility}
 and both reduce to the identity matrix when the non-equilibrium parameter $\tau/\gamma \to 0$.


\section{Kinetic approach: Mesoscopic Virial Equation}
\label{MVE}
In this section, we shall use a  method inspired by the recent paper
by Falasco et al \cite{falasco2015mesoscopic}, who derived a virial equation for non-equilibrium  systems  described by
a Fokker-Planck equation.  
Clausius's virial theorem  follows from the fact that the average of the derivative of a bounded function is zero.
In the specific case of the UCNA,  the result follows when we consider the derivative of an operator, $ \OO$ such that
it  satisfies the following  set of differential equations:
\be
\frac{\partial \OO(\rr_1,\dots,\rr_N) }{\partial r_{\alpha i}}  =\sum_{\delta l}\Gamma_{\alpha i,\delta l} \,r_{\delta l}
\ee
A representation of  $\OO$, a part an arbitrary and inessential constant, is:
\be
\OO =\sum_{\alpha i}        \Bigl(\frac{r_{\alpha i}^2}{2}+\frac{\tau}{\gamma} \frac{\partial\UU_{total}}{\partial r_{\alpha i}} r_{\alpha i} \Bigr)-\frac{\tau}{\gamma} \UU_{total}  
\label{operator}
\ee    
Using  \eqref{operator} and  \eqref{oevolution} explicitly we obtain:
\be
\frac{d \langle\OO\rangle}{dt}=  \int d^{N} \rr P_N (\rr_1,\dots,\rr_N)
 \sum_{\beta k}  \Bigl(   \frac{1}{\gamma } F_{\beta k}    \,r_{\beta k}
+ D     \Gamma^{-1}_{\beta k,\beta k}    \Bigr)
\label{oevolution5}
\ee
 the sums are over all the particles, with locations denoted by $\rr_i$,  on which act the forces $F_i$. Since the particles are bounded by the finite volume V the time average of the l.h.s.
	will approach zero and we can write by splitting the forces into the external, or confining wall forces, and the inter-particle force:
\be
\sum_i^N  \langle (\F^{ext}_i+ \F^{int}_i) \cdot \rr_i \rangle+D\gamma \sum_{\alpha i} \langle   \Gamma^{-1}_{\alpha i,\alpha i}  \rangle=0
\label{virialbaiesi}
\ee
 In the last term the forces are separated in two parts:  wall  and  interparticle  forces. The forces exerted by the bounding walls of the container are macroscopically described as external
pressure: each element $d{\bf A}$  of their area  exerts a force
$-p_v d{\bf A}$ (the subscript V stands for virial)  so that 
$$
\sum_i^N  \langle \F^{ext}_i \cdot \rr_i \rangle =-p_v \oint \rr\cdot d{\bf A}=-p_V V d   \,,
$$
 where $\rr$ is the position vector of the surface element and the last equality follows from the application of the divergence theorem. The internal virial is:
\be
\sum_i^N  \langle \F^{int}_i \cdot \rr_i \rangle=\frac{1}{2}\sum_i\sum_j'  \langle \F_{ij} \cdot (\rr_i-\rr_j) \rangle
 \ee
 where we have symmetrized the sum.
Finally, by approximating the  inverse matrix   $\Gamma^{-1}$  using the methods of appendix \ref{bgamma}  we evaluate the trace of the  $dN\times dN$ matrix by  the following formula:
\be
T_s \sum_i^N \sum_{\alpha }^d   \langle   \Gamma^{-1}_{\alpha i,\alpha i}  \rangle \approx  T_s  \sum_\alpha^d \int d\rr \, \tilde  \Gamma^{-1}_{\alpha \alpha }(\rr)\rho(\rr)
\ee
where in the second equality we have used \eqref{sedici}. We, now, write 
\be
 p_v= \frac{T_s}{d V}   \sum_\alpha^d \int d\rr \, \tilde  \Gamma^{-1}_{\alpha \alpha }(\rr)\rho(\rr)
-\frac{1}{2 d V }   \sum_\alpha^d \int d\rr \,  \int d\rr'   \rho^{(2)}(\rr,\rr')   \frac{ \partial w(\rr-\rr')}{\partial r_\alpha} (r_\alpha-r_\alpha') 
\label{pressurevirial}
\ee
where the second term in \eqref{pressurevirial}
is analogous to the direct contribution to the pressure in passive fluids  stemming from interactions.
Using the approximation \eqref{smallGamma} we obtain the explicit representation
\be
  \frac{1}{d V} T_s  \sum_\alpha^d \int d\rr \, \tilde  \Gamma^{-1}_{\alpha \alpha }(\rr)\rho(\rr)\approx
   \frac{T_s}{V} \Bigl[ N -    \frac{1}{d}\frac{\tau}{\gamma}    
  \int d\rr \sum_\alpha u_{\alpha\alpha}(\rr)\rho(\rr)
    -  \frac{1}{d }\frac{\tau}{\gamma}  \int d\rr \rho(\rr) \int   d\rr'    \rho(\rr') g(\rr,\rr') \sum_\alpha w_{\alpha\alpha}(\rr-\rr')  \Bigl]
 \label{pressureswim}
    \ee
where  the double subscripts indicate second partial derivatives of the potentials  with respect to the coordinate $r_\alpha$.
The first term in the r.h.s.  of \eqref{pressurevirial} contains the ideal gas pressure, $T_t N/V$,   stemming from the translational
degrees of freedom, the swim pressure,  $T_a N/V$, (recall that $T_s=T_t+T_a$) due to the rotational degrees of freedom  and the so-called indirect interaction contribution represented by the second and third term in formula \eqref{pressureswim},   which takes into account the slowing down of active fluids near a boundary or in regions of high density. The indirect interaction pressure involves the interplay between the rotational degrees of freedom and the 
interparticle forces and is a non-equilibrium effect.
In fact, in the limit of $\tau\to 0$  the quantity~\eqref{pressureswim},  reduces to $T_t N/V$, the ideal gas contribution to the pressure.
Expressions for the pressure equivalent to \eqref{pressurevirial}  will be derived by two different approaches in the next sections.
  \section{Statistical mechanical approach}
  \label{stretching}
  The mechanical interpretation of the pressure is not the only one: the pressure and the surface tension
  are also thermodynamic and statistical observables characterising the macrostate of the system. 
   In this section and in the related 
  appendices \ref{volumerescaling} and \ref{arearescaling}
 we derive their expressions by differentiation of the partition function, $Z_N$, with respect to the volume and area 
 and  verify that these derivatives are equal (within the order allowed by the approximations involved) to the 
 MVE expression. In other words the
 thermodynamic-like
 relationship $p_t=-\frac{\partial \Fu}{\partial V}_{T,N}$, where $\Fu$ is defined as $\Fu=-T_s \ln Z_N$
 agrees with $p_v$. As far as the surface tension is concerned the agreement with a  method based on the distribution functions is shown in  section \ref{bauslovett}.
  \subsection{Statistical pressure}
 The logarithmic derivative with respect to the volume is performed employing a volume scaling procedure as illustrated in appendix \ref{volumerescaling}
 with the following result:
 

\be
 p_t  =  \frac{T_s}{3 V}
 \sum_{\alpha }^d \sum_i^N  \langle   \Gamma^{-1}_{\alpha i,\alpha i}  \rangle+  \frac{1}{6 V}\sum_i\sum_j'  \langle \F_{ij} \cdot (\rr_i-\rr_j) \rangle 
\label{virialbaiesiold}
\ee
which is the same expression as eq. \eqref{pressurevirial} obtained by using the kinetic method  . Moreover, we shall demonstrate that
such a definition of pressure coincides (up to order $\tau/\gamma$) with the pressure obtained  by the microscopic 
condition of hydrostatic equilibrium.
Again we can see that the pressure is made up of three contributions:  ideal gas and indirect  pressure both contained in the first   term, as can be seen using eq.\eqref{sedici}, and a direct interparticle interaction  pressure term  contained in the last term.

  \subsection{Statistical surface tension}

 Buff obtained a formula for the surface tension of equilibrium systems by generalizing the stretching method
 of the previous section \cite{buff1952some}.
 We prove that the surface tension obtained through the mechanical definition also satisfies a macroscopic thermodynamic-like
 relationship $\gamma=-T_s\frac{\partial \ln Z_N}{\partial A}_{T,N}$.
 Performing the derivative of the configurational integral $Z_N$ requires  taking into account that the limits of the integrals defining
 $Z_N$ depend on the system area, $A$.  A simple  calculation reported in appendix \ref{arearescaling}  gives:
\be 
\gamma = \frac{1}{ A}  \int_V d\rr_1\dots d\rr_N   P_N( \rr_1,\dots, \rr_N)   \sum_i  \Bigl\{ \Bigl[ x_i \frac{\partial U_{int} }{\partial  x_i}  -z_i \frac{\partial U_{int} }{\partial  z_i} \Bigr] 
+ T_s \frac{\tau}{\gamma} \Bigl[ 
  \frac{\partial^2  \UU_{int}}{\partial x_ i^2 }  -    \frac{\partial^2  \UU_{int}} {\partial z_ i^2 } \Bigr]   \Bigr\}
\label{surfacetension1}
\ee
Since $\UU_{int}$ is a sum of pair potentials using standard manipulations and eq.  \eqref{smallGamma}
 we can rewrite the first contribution  in the integrand, that we identify with the passive surface
tension and obtain:
\be
\gamma =   \frac{1}{ 2 A}  \iint_V d\rr d\rr' \rho(\rr)\rho(\rr') g(\rr,\rr')  \Bigl[ \frac{Y^2}{R} -  \frac{Z^2}{R}  \Bigr]   \frac{\partial}{\partial R }  w( R)+     T_s  \frac{1}{  A}  \int_V d\rr  \rho(\rr)\Bigl[ \tilde \Gamma^{-1}_{zz}(\rr)- \tilde \Gamma^{-1}_{xx}(\rr) \Bigr]  
\label{surface_stretching}
  \ee
with ${\bf R}=(X,Y,Z)=(x-x',y-y',z-z')$. In the last term, we have used the expansion of the matrix $\Gamma$ up to order $\tau/\gamma$ to rewrite the formula
for the surface tension. We shall prove below in section \ref{planar}
 that the surface tension computed by the stretching method gives the same value as the one computed 
by using the concept of work necessary to stretch the interface.

To zeroth order in the parameter $\tau$ the Kirkwood-Buff formula \cite{kirkwood1949statistical} for the surface tension of a planar interface  is recovered, whereas the second term represents the correction due to the activity and is similar to the surface tension found by  Bialk{\'e} et al.  \cite{bialke2015negative}.
Notice that in a passive fluid the last term vanishes, because the $\tilde \Gamma$ matrix becomes the identity matrix.

 \section{Distribution functions approach}
 \label{bauslovett}
 We now develop a method that allows computing the pressure tensor and the surface tension in 
 situations where the fluid can also be inhomogeneous. The starting point is different from the one of the previous section which 
 uses  the Fokker-Planck equation to estimate the virial as in eq. \eqref{virialbaiesi}. We shall use instead the fact that 
 when the system is perturbed starting from an initial steady non-uniform steady state the work done on the system to produce such a result
 can be written in two different ways: either in terms of the work done
  by external forces when the particles are displaced by an infinitesimal amount with respect to their
 initial positions or expressing the work necessary to produce the same deformation ( with respect to the given reference state)  in terms of the product of the stress and strain tensors. Whereas in a uniform simple fluid one needs a single scalar quantity such as the relative volume variation to measure its deformation, in a non-uniform fluid the deformation with respect to a reference state is measured by a displacement vector $\delta {\bf s}(\rr)$ \cite{mistura1985pressure,lautrup2011physics}.
 This method is a generalization to active particles of an approach introduced in the literature of passive fluids
 by Baus and Lovett. Such a method allows defining   the pressure tensor in terms of microscopic molecular distributions
 with the help of the information contained in the first BGY equation \eqref{tssum}.

 Once we endow the first BGY equation with a suitable closure, in practice  expressing the higher correlations in terms of the density profile and of the pair correlation function, it is in principle possible to determine numerically the density profile. 
 On the other hand, even without a numerical solution of the BGY equation we can make predictions about  the relevant
 observables of the problem, such as  the pressure and the surface tension. In equilibrium systems, such quantities are measured unambiguously. Hereafter, we present a mechanical derivation of these two quantities 
 following the method of Baus and Lovett ~\cite{baus1991stress,lovett1991family} ,
and derive the pressure by computing the work necessary to produce an infinitesimal
compression of the system and the surface tension by computing the work necessary to make an
infinitesimal stretching of its surface.
In  the limit $\tau \to 0$ both observables become identical to their equilibrium counterparts.

Following  the book Elasticity Theory  by Landau-Lifshitz   \cite{landau1975elasticity} ) 
the relative variation of volume $\delta V/V$ 
can be expressed in terms of
the strain $
\delta \epsilon_{\alpha\beta}=\frac{1}{2}( \frac{\partial \delta s_\alpha}{\partial r_\beta}+  \frac{\partial \delta s_\beta}{\partial r_\alpha})
$ associated with
a local displacement  $\delta {\bf s(\rr)}$ of the fluid,
such  that matter at $\rr$ is displaced to a new position $\rr'=\rr+\delta{\bf s(\rr)} $, and the relative volume change \cite{schofield1982statistical} is:
$
\frac{\delta V}{V}= \nabla\cdot \delta {\bf s(\rr)} 
$.
 The condition of mechanical  (hydrostatic) equilibrium in the system dictates
that the pressure tensor $p_{\alpha\beta}$ in the absence of currents is subjected to the constraint:
\be
\sum_\beta \frac{\partial}{\partial r_\beta} p_{\alpha\beta} (\rr)=f_{\alpha}(\rr).
\label{twenty3}
\ee
where ${\bf f}(\rr)=\rho(\rr) {\bf F}(\rr)$ is  the external force per unit volume responsible for confining
the system in space and is a body force.
We can express the work done  on the system when each particle $i$ is displaced by an amount $\delta {\bf s}(\rr_i)$
as:
\be
\delta W=\int_V d\rr \sum_\alpha f_{\alpha}(\rr)\delta s_\alpha(\rr)+\oint_A dA \sum_\alpha   t_\alpha(\rr)\delta s_\alpha(\rr)
\label{ventiquattrox}
\ee
where $t_\alpha(\rr)\equiv- p_{\alpha\beta}(\rr) n_\alpha(\rr)$ is the $\alpha$ component of the stress vector acting on surface of  area $A$ of the system at position, and ${\bf n}(\rr)$ the normal to the surface $A$ at $\rr$.
The second term in the r.h.s represents the work
done by the surface stress and notice that the minus sign is a consequence of our convention of not introducing the stress,
but to use directly the pressure tensor which is minus the stress tensor.
Following Landau-Lifshitz  
let us imagine, now, that the system is undeformed at infinity and remove at infinity the surface where the integral is performed
so that the second term in the r.h.s of \eqref{ventiquattrox} vanishes
and only the  first term remains.
On the other hand, the work \eqref{ventiquattrox}  can also be expressed  in terms of the pressure tensor by  using \eqref{twenty3}  :
\be
\delta W=\int_V d\rr \sum_{\alpha\beta} \frac{\partial}{\partial r_\beta} p_{\alpha\beta} (\rr)\delta s_\alpha(\rr)
\label{lafa2}
\ee
As a consequence of
$
\oint_A dA \sum_\alpha   t_\alpha(\rr)\delta s_\alpha(\rr)=0$, we finally rewrite  the work
in terms of the strain and of the pressure tensor, $p_{\alpha\beta}$. 
\be
\delta W=-  \int_V d\rr  \sum_{\alpha\beta}p_{\alpha\beta}(\rr) \delta \epsilon_{\alpha\beta} (\rr)
\label{pstrain}
\ee
 In order to obtain an expression for the bulk pressure in a uniform and isotropic system, 
 where $p_{\alpha\beta}(\rr)=p_V\delta_{\alpha\beta}$,
we, now,  assume the deformation to be a uniform infinitesimal dilatation
$\delta {\bf s(\rr)}=\lambda \rr $, corresponding to a relative volume change 
$
\frac{\delta V}{V}= 3 \lambda
$,
 According to \eqref{pstrain}  the work of deformation is $\delta W= -\lambda \int_V d\rr \sum_\alpha p_{\alpha\alpha}(\rr) =-3 \lambda p_V V$. On the other hand, by computing the work using the body force we  obtain:
$
\delta W= \lambda \int_V d^3 \rr \, \rho(\rr)  {\bf F} (\rr)\cdot \rr ,
$ and by equating these last two expressions  we find the volume averaged  pressure
\be
p_V
=-\frac{1}{3V} \int_V d\rr  \rho(\rr) \F(\rr)    \cdot \rr .
\label{virialpres2} 
\ee
Finally,  with the help of
the BGY equation  \eqref{rho1bgy} we eliminate $\rho(\rr){\bf F}(\rr)$, substitute in \eqref{virialpres2}:
and express the pressure in terms of molecular distributions:
\be
p_V=-\frac{T_s}{3V} \int d\rr \sum_{\alpha\beta} \frac{\partial }{\partial r_{\beta }}\Bigl[ \tilde \Gamma^{-1}_{\alpha\beta} (\rr)\rho(\rr)  \Bigl] r_\alpha
- \frac{1}{3V}\int d\rr  \int   d\rr'   \rho^{(2)}(\rr,\rr')   \sum_\alpha  \frac{ \partial w(\rr-\rr') }{ \partial r_{\alpha}} r_\alpha
\label{ventotto}
\ee
Similarly to eq. \eqref{pressurevirial},
the first term in equation \eqref{ventotto}, that in the following we shall call  "active"  for short, is the sum of three contributions: ideal gas, swim   and indirect contribution. It can be integrated by parts and,  after
discarding  a surface term becomes
\be
p_V^{active}
=\frac{T_s}{3V}\sum_{\alpha}  \int d\rr   \tilde \Gamma^{-1}_{\alpha\alpha} (\rr)\rho(\rr) 
\label{pactive}
\ee
and turns out to have the same form as \eqref{pressureswim}.
The second term in the left hand side of eq. \eqref{ventotto}, instead, represents the so-called direct interaction term due to intermolecular
forces and can be expressed as:
\be
p_V^{direct}= - \frac{1}{6V}  \int  d\rr \int   d\rr'   \rho^{(2)}(\rr,\rr') \sum_\alpha(r_\alpha-r'_{\alpha})  \frac{ \partial w(\rr-\rr') }{ \partial r_{\alpha }} 
\ee
while the total pressure is given by the sum
$p_V=p_V^{active}+p_V^{direct} $ and is the same as the MVE result of \eqref{pressurevirial} and thus to
\eqref{virialbaiesiold}.

The work of deformation method, which yields the same result for the pressure of a uniform system as the MVE approach of section \ref{MVE}, will be now applied hereafter to two different non-uniform situations: a) an active fluid in the presence
of a flat boundary and b) in the presence of an interface between two phases.
\subsection{Pressure against a planar wall}
We represent a wall by means of an external soft-repulsive  potential $u(\rr)$  rapidly decaying away from the wall
and derive the corresponding expression of the pressure tensor.  In this case, we consider an anisotropic deformation, $ \delta {\bf s}(\rr)$, affecting only the direction normal to the wall, and to this purpose
choose a different parametrization.
To include the planar wall we consider a cubic system
\be
0\leq z\leq L_z  \, ;\,\,\,\, 
- \frac{1}{2}L_x\leq x \leq   \frac{1}{2}L_x \, ;\, \,\,\,\,
- \frac{1}{2}L_y\leq y \leq   \frac{1}{2}L_y
\ee
and locate the wall at $z=0$. The displacement field is now represent by
$
\delta{\bf s}(\rr)=(0,0,\lambda(z-\bar z) \theta(z-\bar z))
$
and  the associated volume change reads: 
$
\delta V=\int d\rr\,\nabla\cdot \delta {\bf s(\rr)}= \lambda L_x L_y (L_z-\bar z)
$,
where $z=\bar z$ denotes a reference plane inside the system, where ${\bf f}(\rr)$ is negligible and non zero
 only near the boundaries of the system, so that by  \eqref{twenty3}
$p_{zz}(\rr)$ is nearly constant for $z\geq \bar z$. After the displacement the wall originally at $z=L_z$ is located
at $z=L_z+\lambda(L_z-\bar z)$.
The work done on the system when the volume changes by $\delta V$ is according to \eqref{pstrain}:
\be
\delta W=-  \lambda L_x L_y \int_{\bar z}^{L_z} dz p_{zz}(z) 
\label{pressurelb}
\ee
We again
express ${\bf f}(\rr)$ in terms of distribution functions with the help of 
 \eqref{rho1bgy} and compute the work of deformation using \eqref{lafa2}
\be
\delta W= \int d\rr\, \lambda (z-\bar z)\theta(z-\bar z) \Bigl\{ \sum_{\beta} \frac{\partial }{\partial r_{\beta  }}\Bigl[ T_s \tilde \Gamma^{-1}_{z \beta} (\rr)\rho(\rr)  \Bigl] +
  \int   d\rr'   \rho^{(2)}(\rr,\rr')    \frac{ \partial w(\rr-\rr') }{ \partial z}  \Bigr\}
\label{worklb}
\ee
Equating \eqref{pressurelb} and \eqref{worklb} and differentiating both sides w.r.t. $\bar z$ we find:
\be
p_{zz} (\bar z) L_x L_y =- \int d\rr\, \theta(z-\bar z) \Bigl\{ \sum_{\beta} \frac{\partial }{\partial r_{\beta  }}\Bigl[ T_s \tilde \Gamma^{-1}_{z \beta} (\rr)\rho(\rr)  \Bigl] +
  \int   d\rr'   \rho^{(2)}(\rr,\rr')    \frac{ \partial w(\rr-\rr') }{ \partial z}  \Bigr\}
\ee
Integrating by parts the first term and recalling that the surface terms vanish outside the finite volume of the system we get:
\be
p_{zz}(\bar z)= \int \frac{d\rr}{L_x L_y}\,  \Bigl\{ \delta(z-\bar z)  \, T_s   \tilde \Gamma^{-1}_{zz} (\rr)\rho(z)   -  \theta(z-\bar z) 
  \int   d\rr'   \rho^{(2)}(\rr,\rr')    \frac{ \partial w(\rr-\rr') }{ \partial z}  \Bigr\}
\ee
Finally, we let $L_x\to \infty$ and $L_y\to\infty$   transform the formula (see Baus-Lovett   \cite{baus1991stress}) and obtain the final form:
\be
p_{zz}(\bar z)=  T_s   \tilde \Gamma^{-1}_{zz} (\bar z)\rho(\bar z)     - \int dz 
  \int   d\rr'   \theta(z-\bar z)    \theta(\bar z- z')  \rho^{(2)}(\rr,\rr')    \frac{ z-z' }{ |\rr-\rr'|} w'(|\rr-\rr'|) 
\ee
As discussed by Baus and Lovett the meaning of such an equation is that the pressure $p_{zz}(\bar z)$
can be measured by evaluating the first term at any plane $\bar z$ inside the system and the interaction contribution
due to the direct forces acting across the $\bar z$ plane so that particles at $z$ and at $z'$ are on opposite sides
of the mathematical surface at $\bar z$ (i.e. for $z>\bar z$ and $z'<\bar z$).
An analogous formula has been derived by Solon et al.
\be
p_{zz}(\bar z)=\frac{  T_s \rho(\bar z) }{  \tilde \Gamma_{zz} (\bar z)}    - 2\pi \int_{\bar z}^\infty dz  
 \int_0^{\bar z} dz'   \int_0^\infty dR R  \rho^{(2)}(z,z',R)    \frac{ z-z' }{ \sqrt{(z-z')^2+R^2}} w'(z,z',R) 
\ee

On the other hand if $\bar z \to \infty$ the density $\rho(\bar z)$ becomes uniform and the formula above simplifies:
\be
p=T_s \frac{\rho_b}{  \tilde \Gamma_{b}}-\frac{2}{3}\pi \rho_b^2\int_0^\infty dr r^3 g_b(r)w'(r)
\ee
with
\be
\tilde \Gamma_b=1+ \frac{4}{3} \frac{\tau}{\gamma} \pi \rho_b \int_0^\infty dr r^2 g_b(r)[w''(r) +2 \frac{w'(r)}{r}] 
\ee
where the subscript b denotes the bulk value of the corresponding quantity.

\subsection{Surface tension of a planar interface}
\label{planar}
We are, now, interested in deriving a formula for the surface tension of a planar interface normal to the $z$ direction 
and separating two different phases. 
The surface tension, $\gamma$,  is computed from the mechanical  definition of work done on the system as a consequence of an isothermal 
change, $\delta A$, of the area $A$ of the interface:
\be
\delta W=-p_V\delta V+\gamma\delta A
\ee
In order to compute $\gamma$,
we remark that by symmetry the pressure tensor is diagonal, depends only on the $z$ coordinate and can be represented  
by only two functions: the normal component $p_N(z)=p_{zz}(z)$ and the tangential component $p_T(z)=p_{xx}(z)=p_{yy}(z)$, which are not equal as in the bulk
because the latter includes the tension of the interface.
 Given a rectangular cuboid of base area $A$ and height $L$, we follow a simple argument by Rowlinson and Widom  
\cite{rowlinson2013molecular} to compute
 the work necessary to increase  by an amount $\delta A$ the area  without changing the box volume: it 
 is the sum of two contributions, 
 a) the work necessary to keep the volume constant by applying  during this process a normal pressure $p_N$ to the stretched interface 
 and b) the tangential work to stretch the interface. Assuming that $p_N$ does not change with $z$ due to the condition of hydrostatic equilibrium,  we write:
 $$
 \delta W= -p_N A \delta L   -\delta A\int _{-L/2}^{L/2}  dz p_T(z) 
 $$
and since the volume variation must be zero, to linear order we have  $\delta V=A\delta L+L\delta A=0$, we conclude that $\delta L=-L \delta A/A$ and
 \be
 \delta W=\delta A\int _{-L/2}^{L/2}  dz [p_N-p_T(z)] ,
 \label{deltaW}
 \ee
 and identify $\gamma$ with the integral in \eqref{deltaW}.
 Using the strain-stress formalism we can derive the same result and compute explicitly the pressure components by considering
  the work done
by the internal forces:
\bea
\delta W &=&   - \int_V d^3\rr[ p_N(z) \delta \epsilon_{zz}+ p_T(z) \delta \epsilon_{xx}+p_T(z) \delta \epsilon_{yy}]
\nonumber\\ &=&
-  \frac{1}{3}  \int_V d^3\rr \,\Bigl( (p_N(z)+2 p_T(z)) \ [\delta \epsilon_{zz}+  \delta \epsilon_{xx}+\delta \epsilon_{yy}]-(p_N(z)-p_T(z)) [  \delta \epsilon_{xx}+ \delta \epsilon_{yy}- 2 \delta \epsilon_{zz}]  \Bigr).
\label{quaranta3}
\eea
The first term represents the trace of the pressure tensor and provides average pressure $p_{av}(z)= (p_N(z)+2 p_T(z))/3$  of the non-uniform system.
Let us now assume a displacement of the form:
 $\delta {\bf s(\rr)}=\lambda (x,0,-z) $ corresponding to an isochoric process, i.e. $\nabla\cdot \delta {\bf s(\rr)}=0$,
and to a relative area change 
$\frac{\delta A}{A}=\lambda$  of the surface normal to the z-direction, so that  \eqref{quaranta3} can be written as:
\be
\delta W=   \lambda  \int_V d^3\rr (p_N(z)-p_T(z)) = \delta A \int d z (p_N(z)-p_T(z)) =\gamma \delta A .
\label{pn_pt}
\ee
We, now, perform the calculations of the work done by  the external force
( $\delta W=\int_V d\rr {\bf f}\cdot \delta {\bf s}(\rr)$)
and equate it to  $\delta W=\gamma\delta A$. We obtain  the following expression for the surface tension:
\be
\gamma=\frac{1}{A} \int_V d\rr [xf_x(\rr)-zf_z(\rr)] 
\label{gammaforce}
\ee
and proceed to  eliminate the components of ${\bf f}$ in favor  of the microscopic distribution functions in  \eqref{gammaforce} using \eqref{rho1bgy}. The resulting expression
of the surface tension can be divided in two pieces as  $\gamma=\gamma^{active}+\gamma^{direct}$: the first
one contains ideal, swim and indirect contributions and after integrating by parts reads: 
\be
\gamma^{active}=\frac{T_s}{A} \int_V d\rr     \sum_\beta \frac{\partial }{\partial r_{\beta }}    [ x \tilde \Gamma^{-1}_{x\beta} (\rr)\rho(\rr)
- z\tilde \Gamma^{-1}_{z\beta} (\rr)\rho(\rr)]-
\frac{T_s}{A} \int_V d\rr     \sum_\beta   [ \delta_{\beta x} \tilde \Gamma^{-1}_{x\beta}(\rr)
- \delta_{\beta z} \tilde \Gamma^{-1}_{z\beta} (\rr) ]\rho(\rr) .
\ee
Taking into account that  the first integral in the r.h.s. vanishes  we obtain:
\be
\gamma^{active}=\frac{T_s}{A} \int_V d\rr       [  \tilde \Gamma^{-1}_{zz} (\rr)
- \tilde \Gamma^{-1}_{xx} (\rr)] \rho(\rr) .
\label{gammaactive}
\ee
The second contribution to the surface tension, stemming from  direct interactions  and represented by  the last term in the r.h.s. of \eqref{rho1bgy},  is a standard calculation in equilibrium fluids and results in the following expression:
\be
\gamma^{direct}=  \frac{1}{2 A} \int d\rr\int   d\rr'   \rho^{(2)}(\rr,\rr') \Bigl(\frac{(x-x')^2  -(z-z')^2 }{|\rr-\rr'|} \Bigr) \frac{ \partial w(\rr-\rr') }{ \partial r}  
\label{gammapassive}
\ee
By putting together \eqref{gammaactive}  and  \eqref{gammapassive}  we see that this prescription for the surface tension gives the same result up to order $\tau/\gamma$ as formula\eqref{surface_stretching} which was obtained by an isochoric deformation of the system and coordinate rescaling.

In order to evaluate $\gamma^{active}$ we need the explicit form of
the diagonal $\tilde \Gamma_{\alpha\beta}$ matrix elements:
\be
 \tilde \Gamma_{N(T)}(z)  =      1+\frac{\tau}{\gamma} \int d\rr' w_{zz(xx)}(\rr,\rr')\rho(z') g(\rr,\rr')  .
 \label{GammaNT}
 \ee
Using \eqref{gammaactive} and \eqref{gammapassive} and the second equality in \eqref{pn_pt}
 we may identify the active and passive part of the normal and tangential pressure tensor as:
\be
\Bigl(\begin{array}{cc}
 p_N^{active}(z)   \\      p_T^{active}(z) \end{array} \Bigr) 
=T_s  \rho(z)  \Bigl(\begin{array}{cc}
   \frac{1 }{\Gamma_N(z)} \\      \frac{1 }{\Gamma_T(z)} \end{array} \Bigr) 
\ee

and
\be
\Bigl(\begin{array}{cc}
 p_N^{direct}(z)   \\      p_T^{direct}(z) \end{array} \Bigr) 
= -\frac{1}{2} \rho(z)\int   d\rr'   g(\rr,\rr') \rho(z') \Bigl(\begin{array}{cc}
   (z-z')^2 \\      ( x-x')^2  \end{array} \Bigr) \frac{1 }{|\rr-\rr'|} \frac{ \partial w(\rr-\rr') }{ \partial r}  
\ee

Notice that the active pressure is higher in the low-density region than in the dense phase
because of the mobility reduction \cite{bialke2014negative}.

Summing the two expressions we obtain the total surface tension
$
p_{N(T)}= p_{N(T)}^{active}(z)+  p_{N(T)}^{direct}(z) $.
Notice that the scalar pressure defined as:
$
p_{av}(z)=\frac{1}{3} Tr \, p_{\alpha\beta}(z)
$
is not constant through the interface, whereas $p_N(z)$ must be constant in order to ensure mechanical equilibrium
in the absence of external stabilizing fields.

To conclude, we have seen that the activity modifies the ideal gas term of the  pressure
by the presence of the friction matrix $\tilde \Gamma_{\alpha\beta}(\rr)$, besides modifying the density distribution and the pair correlation function.
This trend has been observed by the three methods considered 
 in all the scrutinized quantities, namely the surface tension and the components of the pressure tensor.
\subsection{Comment about the swim pressure of Solon et al.}

In order to make contact with the approach proposed by Solon et al.  \cite{solon2014pressure,solon2014pressure2}  we compare our result
~\eqref{pactive} with their expression for the swim pressure which in the present notation reads:
$$
 p^{swim}=\frac{\rho}{d} \tau \gamma v_0 v(\rho)
$$
where $v(\rho)$  and $v_0$ are  the average speeds  of a particle along its direction of propulsion with and without 
 interparticle interactions, respectively.  In the interacting case,  these authors express the density dependent velocity as  $v(\rho)=(v_0+I_2/\rho)$ 
 where $I_2$ is expressed in terms of the pair potential.
Separating the swim pressure  and identifying
$T_a/\gamma=D_a=v_0^2 \tau/d$ we have
\be
p^{swim}=\rho\frac{T_a}{d} \tau \gamma v_0( v_0+ (v(\rho)-v_0))=  \rho T_a + \frac{\rho}{d} \tau \gamma v_0 (v(\rho)-v_0) . 
\label{pressuresolon}
\ee
Such a result can be compared with our expression for the active pressure ~\eqref{pactive} in the case where, for the sake of simplicity we set $T_t=0$ and assume a uniform density so that
$p^{active}=T_a \frac{ \rho}{d}   \sum_{\alpha=1}^d   \tilde \Gamma^{-1}_{\alpha\alpha} $.
By using a result recently derived \cite{marconi2015velocity}. we relate the variance of the velocity of the particles in the UCNA model to the inverse friction matrix $\tilde \Gamma$ according to
$ \sum_{\alpha=1}^d   \langle v_{\alpha}v_{\alpha}\rangle 
= \frac{D_a}{\tau}   \sum_{\alpha=1}^d   \tilde \Gamma^{-1}_{\alpha\alpha}
$
and rewrite the active pressure as:
\be
 p^{active}=\rho  \frac{T_a}{d}   \sum_{\alpha=1}^d [1+  (\tilde \Gamma^{-1}_{\alpha\alpha}-1)]
\approx \rho T_a   +  \frac{\rho}{d} \tau \gamma[ \langle v_\alpha v_\alpha \rangle- \langle v^0_\alpha v^0_\alpha \rangle] .
 \label{pact}
\ee
with $ \langle v^0_\alpha v^0_\alpha \rangle]\equiv D_a/\tau$. One can observe the striking similarity between  eqs. \eqref{pressuresolon}  and \eqref{pact},
in both  appears the difference between the effective velocity  $v$   and the velocity of an isolated particle,  
so that in both equations the excess term vanishes when $\rho\to 0$.

 \section{conclusions} 
 \label{conclusions} 
  We have studied different routes to pressure for a particular model system 
of active particles.
We have focused on an argument which has recently generated some debate: whether the pressure in active fluids is a state function or not. By an analysis of models with anisotropic interactions or with quorum sensing, it has been reported that  this is generally not the case \cite{solon2014pressure,solon2014pressure2}.  Differently for our model we have found that, within the limits of our approximations, the pressure of an assembly of active particles with repulsive spherical interactions is a state function and can be computed from a partition function  and this coincides with other definitions of pressure obtained from a mechanical perspective.  
Very recently, Speck \cite{tspeck2016} using a stochastic thermodynamic approach has derived expressions for the pressure and the surface  tension of active Brownian particles that agree with ours.

Summarizing we have computed the pressure and the surface tension of an active system within the UCNA
and found that the results of different methods are in agreement { as long as we limit ourselves to first order in the relaxation time of the active noise}.
Moreover, all the quantities reduce smoothly to their equilibrium values when this non-equilibrium parameter
vanishes. In a forthcoming publication, we shall also show that for a simple model system of  harmonic dumbbells these theoretical predictions are exact and the agreement between different methods is valid to any order.  Our results also suggest to compare the pressure obtained numerically in other simple models of active particles (e.g. in active Brownian and run and tumble particles) and check if the agreement among different definitions of pressure is preserved when the relaxation time of the noise is small.

\section*{Acknowledgments}
We thank Massimo Cencini for a very constructive exchange of opinions 
and M. Dijkstra, N.Gnan, A. Puglisi  and R. Di Leonardo for  discussions.
C. Maggi acknowledges support from the European Research Council under the European Union's Seventh Framework programme
(FP7/2007-2013)/ERC Grant agreement  No. 307940.

\bibliography{pressureactive.bib}

 \appendix
 
  
\section{Derivation of the mobility term in the BGY equation }
\label{offdiagonal}

The stationary distribution $P_N(\rr_1,\dots,\rr_N)$ 
obeys a first order partial differential equation which relates
the probability distribution to the potential  $\UU_{total}$   and from this equation,  one obtains  equations for
the
marginalized distribution functions $P^{(n)}_N(\rr_1,\dots,\rr_n)=\int d\rr_{n+1}\dots d\rr_N P_N(\rr_1,\dots ,\rr_N )$
of different orders (see ref. \cite{marconi2015towards}). In particular, one obtains:
\bea
&&
  T_s \int \int d\rr_2\dots d\rr_N \sum_\beta \sum_k \frac{\partial }{\partial r_{\beta k}}[  \Gamma^{-1}_{\alpha 1,\beta k}(\rr_1,\dots,\rr_N) P_N(\rr_1,\dots,\rr_N)]=    
-  P^{(1)}_N(\rr_1)  \frac{\partial u(\rr_{ 1})}{\partial r_{\alpha 1}} 
   \nonumber\\
    && 
  - (N-1) \int d\rr_2   P^{(2)}_N(\rr_1,\rr_2)   \frac{ \partial w(\rr_1-\rr_2) }{ \partial r_{\alpha 1}}   .
\label{p2distrb}
\eea

We write the l.h.s  of \eqref{p2distrb} as the sum of two contributions $D_{\alpha 1,\beta 1}(\rr_1)$ and $\sum_{\beta,k} E_{\alpha 1,\beta k}(\rr_1)$, diagonal (D) and off-diagonal (E) terms, respectively :
\be
D_{\alpha 1,\beta 1}(\rr_1)\equiv  \frac{\partial}{\partial r_{\alpha 1}} 
 T_s \int \int d\rr_2\dots d\rr_N \Gamma^{-1}_{\alpha 1,\beta 1}(\rr_1,\dots,\rr_N) P_N(\rr_1,\dots,\rr_N)
 \label{matrixinversea} 
\ee
and
\be
\sum_\beta \sum_{k\neq 1} E_{\alpha 1,\beta k}(\rr_1)\equiv
  T_s \int \int d\rr_2\dots d\rr_N \sum_\beta \sum_{k\neq 1} \frac{\partial }{\partial r_{\beta k}}[  \Gamma^{-1}_{\alpha 1,\beta k}(\rr_1,\dots,\rr_N) P_N(\rr_1,\dots,\rr_N)]
  \ee
In the limit of $\tau/\gamma$ small  we may approximate the inverse matrix $\Gamma$ as
\be
\Gamma^{-1}_{\alpha i,\beta k} \approx
\Bigl( \delta_{\alpha\beta}-\frac{\tau}{\gamma} u_{\alpha\beta}(\rr_i)-
\frac{\tau}{\gamma}  \sum_j w_{\alpha\beta}(\rr_i-\rr_j) \Bigl)\delta_{ik}
+\frac{\tau}{\gamma}   w_{\alpha\beta}(\rr_i-\rr_k)
(1-\delta_{ik}) .
\label{matrixinverse2} 
\ee
where $u_{\alpha\beta}(\rr)\equiv \frac{\partial^2 u(\rr)}{\partial r_\alpha \partial r_\beta}$ and
$w_{\alpha\beta}(\rr)\equiv \frac{\partial^2 w(\rr)}{\partial r_\alpha \partial r_\beta}$.
Substituting we obtain the diagonal term
\be
D_{\alpha 1,\beta 1}(\rr_1)\approx
T_s\frac{\partial}{\partial r_{\alpha 1}}  \Bigl( P_1(\rr_1)[ \delta_{\alpha\beta}-\frac{\tau}{\gamma} u_{\alpha\beta}(\rr_1)]-
\frac{\tau}{\gamma} (N-1)  \int d\rr'  P_2(\rr_1,\rr') w_{\alpha\beta}(\rr_1-\rr')\Bigr)
\label{delement}
\ee
whereas 
the individual contributions from the  off diagonal elements ($k\neq 1$) are:
\be
\sum_\beta E_{\alpha 1,\beta k}(\rr_1)=
T_s \frac{\tau}{\gamma} \int d\qq_k   \sum_\beta  \frac{\partial}{\partial q_{\beta k}} \Bigl(P_2(\rr_1,\qq_k)  w_{\alpha\beta}(\rr_1-\qq_k)   \Bigl)
\ee

By integrating with respect to $q_{\beta k}$ we find
\be
\sum_\beta E_{\alpha 1,\beta k}(\rr_1)\approx 
 \frac{\tau}{\gamma} \int dq_{xk} dq_{yk} dq_{zk}  \Bigl(P_2(\rr_1,\qq_k)   \sum_\beta w_{\alpha\beta}(\rr_1-\qq_k)   \Bigl) [\delta(q_{\beta k}-L^+_{\beta}) -\delta(q_{\beta k}-L^-_{\beta})]
\label{eelement}
\ee
where $L^{\pm}_{\beta}$ are the coordinates of the boundaries of the system in the $\beta$ direction.
We discard such a boundary term   , i.e. we set  $\sum_\beta E_{\alpha 1,\beta k}(\rr_1)=0$, and 
using  \eqref{p2distrb} and \eqref{matrixinverse2}, \eqref{delement} , \eqref{eelement}   obtain
 the result
\eqref{tssum} expressed in terms of density distributions.

\section{Approximate form of the inverse matrix $\bGamma^{-1}$}
\label{bgamma}
Using eq. \eqref{matrixinverse2}  we compute  the average of the trace of the inverse matrix $\Gamma$ which 
can be written as:
 \be
\sum_i^N \sum_{\alpha }^d   \langle   \Gamma^{-1}_{\alpha i,\alpha i} \rangle \approx
 N \sum_{\alpha }^d  \int d\rr \Bigl( P_1(\rr) [ 1-\frac{\tau}{\gamma} u_{\alpha\alpha}(\rr)]-
\frac{\tau}{\gamma} (N-1)  \int d\rr'  P_2(\rr,\rr') w_{\alpha\alpha}(\rr-\rr')\Bigr)
\label{sedici}
\ee
and switching to density variables $\rho(\rr)=N P_N^{(1)}(\rr)$  and $\rho^{(2)}(\rr,\rr')=\rho(\rr)\rho(\rr') g(\rr,\rr')= N(N-1)
 P_N^{(2)}(\rr,\rr')$ , the one- and two-particle distribution functions, respectively we find:
 \be
 \sum_i^N \sum_{\alpha }^d   \langle   \Gamma^{-1}_{\alpha i,\alpha i} \rangle \approx \int d\rr \rho(\rr)  \tilde \Gamma^{-1}_{\alpha \alpha } (\rr)
 \ee
where 
 $\tilde \Gamma$ is  the $d\times d$ matrix defined  by eq.\eqref{smallGamma}.

\section{Evolution of operators} 
\label{MVEappendix}

The time evolution of the system is given by the Fokker-Planck equation  which reads (see ref. \cite{marconi2015towards}):
\bea
\frac{ \partial P_N(\rr_1,\dots,\rr_N;t)}{\partial t}=-\sum_{\alpha i}  \frac{\partial }{\partial r_{\alpha i}} \sum_{\beta k} \Gamma^{-1}_{\alpha i,\beta k} \Bigl(\frac{1}{\gamma } F_{\beta k}  P_N 
- D \sum_{\gamma j}      \frac{\partial }{\partial r_{\gamma j}}
 [ \Gamma^{-1}_{\gamma j ,\beta k}   P_N ] \Bigr)
\label{FPE}
\eea
where Greek indexes stand for Cartesian components. Consequently the evolution equation of 
the statistical average $\langle \OO(t)\rangle\equiv  \int d^{N} \rr P_N (\rr_1,\dots,\rr_N;t) \OO(\rr_1,\dots,\rr_N)$ of an arbitrary operator 
$\OO$ of the variables $\rr_i$ is given by
\be
\frac{d \langle\OO\rangle}{dt}=  \int d^{N} \rr P_N (\rr_1,\dots,\rr_N;t)
\sum_{\alpha i}  \sum_{\beta k} \Gamma^{-1}_{\alpha i,\beta k} \Bigl(\frac{1}{\gamma } F_{\beta k}    \frac{\partial  \OO}{\partial r_{\alpha i}} 
+ D   \frac{\partial }{\partial r_{\alpha i}}  \sum_{\gamma j}      
 \Gamma^{-1}_{\gamma j ,\beta k} \frac{\partial \OO }{\partial r_{\gamma j}}    \Bigr)
\label{oevolution}
\ee


 \section{Derivative of effective free energy with respect to the enclosing volume}
 \label{volumerescaling}
 Performing the derivative of the configurational integral $Z_N$ requires  taking into account that the limits of the integrals defining
 $Z_N$ depend on the system volume, $V$. To properly take the derivative we non-dimensionalize each particle
 coordinate by the size of the box, by making the change $\rr_i\to L {\bf q}_i$,
where the $-\infty \leq q_{i\alpha} \leq \infty$.
We  evaluate the derivative 
$p_t=T_s\frac{\partial \ln Z_N}{\partial V}$.
and show that up to first order in the perturbation parameter $\tau/\gamma$ this quantity 
coincides with the mechanical pressure and with the virial pressure.
We assume that $\UU_{total}$ includes only pairwise forces and a confining external potential with the following scaling on the linear
size of the system, $L$:
\be
\UU_{total}=\UU_{int}(\rr_1,\dots,\rr_N) + U_{wall}(\frac{\rr_1}{L},\dots ,\frac{\rr_N}{L})
\ee
$U_{wall}$ is a smooth function becoming infinite when $r_\alpha\to \pm \infty$, in particular if $|r_\alpha|>L$ so that the density
becomes exponentially small. 
The volume dependence of $Z_N$ is set by the  typical volume $V=L^3$ since the domain of integration is infinite but
$L$ is the typical size of the container.
In order to obtain the formula \eqref{virialbaiesiold}   we rescale the coordinates
$\rr_i\to L {\bf q}_i$.
  \be
   Z_N  =L^{3 N}   \int_{-\infty}^\infty   d\qq_1\dots  \int_{-\infty}^\infty d \qq_N \exp \Bigl(- \frac{\Ha( L \qq_1,\dots, L \qq_N)}{T_s} \Bigr)
   \label{probabilitydistr}
    \ee
    and evaluate the derivative
    \be
  \frac{\partial  Z_N}{\partial V}=  \frac{N}{V} Z_N-   V^N \frac{1}{T_s}
   \int_{-\infty}^\infty   d\qq_1\dots  \int_{-\infty}^\infty d \qq_N 
   \exp \Bigl(- \frac{\Ha( L \qq_1,\dots,  L\qq_N)}{T_s} \Bigr)  \frac{\partial \Ha( L \qq_1,\dots, L\qq_N)}{\partial  L }  
\ee
As a consequence of the scaling assumed for the wall potential, the derivative of the wall potential drops,
because $   U_{wall}(\frac{\rr_1}{L},\dots ,\frac{\rr_N}{L})  = U_{wall}( \qq_1,\dots, \qq_N)$ does not contain any $L$ dependence and we obtain:
  \be
  \frac{\partial  Z_N}{\partial V}=  \frac{N}{V} Z_N- \frac{1}{3 T_s}  V^N \frac{L}{V}    \int_{-\infty}^\infty   d\qq_1\dots  \int_{-\infty}^\infty d \qq_N 
   \exp \Bigl(- \frac{\Ha( L \qq_1,\dots,  L\qq_N)}{T_s} \Bigr) 
  \sum_i  \qq_i \cdot  \frac{\partial \Ha_1( L \qq_1,\dots, L\qq_N)}{\partial  ( L\qq_i)}  
\ee
where now:
\be 
\Ha_1(\rr_1,\dots,\rr_N) = \UU_{int}(\rr_1,\dots,\rr_N)  +\frac{\tau}{2 \gamma}
    \sum_k^N \Bigr(\frac{\partial \UU_{total}(\rr_1,\dots,\rr_N)}{\partial \rr_k} \Bigl)^2 
    - T_s  \ln  |\det ( I+ \frac{\tau}{\gamma}  \nabla \nabla \UU_{total}(\rr_1,\dots,\rr_N)  |
\ee
Notice that the wall potential  $W$ is absent in the linear term.
We restore, now, the  original dimensional variables in the integrals:
   \be
 p_t=T_s\frac{1}{Z_N} \frac{\partial  Z_N}{\partial V}= T_s \frac{N}{V} - \frac{1}{3 V}  \frac{1}{Z_N} \int   d\rr_1\dots \int  d \rr_N \exp \Bigl(- \frac{\Ha( \rr_1,\dots, \rr_N)}{T_s} \Bigr) \sum_i  \rr_i \cdot  \frac{\partial \Ha_1( \rr_1,\dots, \rr_N)}{\partial  \rr_i}  
\ee
In analogy with equilibrium statistical mechanics and by dimensional considerations, we have identified
the volume derivative of the logarithm of the partition function with a pressure, $p_t$, which reads:
\be
p_t=\frac{N T_s}{V}-\frac{1}{3 V} \langle\sum_i \rr_i \cdot \ \frac{\partial \UU_{int}}{\partial \rr_i} \rangle -\frac{1}{3 V} 
\langle     \sum_i  \rr_i \cdot  \frac{\partial }{\partial  \rr_i} \Bigl[      \frac{\tau}{2 \gamma}
    \sum_k^N \Bigr(\frac{\partial \UU_{total}(\rr_1,\dots,\rr_N)}{\partial \rr_k} \Bigl)^2 
    - T_s \ln     |\det ( I+ \frac{\tau}{\gamma}  \nabla \nabla \UU_{total}(\rr_1,\dots,\rr_N)  | \Bigr] \rangle .
    \ee
    
    {\it Proof of the equivalence with the virial pressure}:
 in order to prove that $p_t$ has a physical meaning, we must ascertain whether it
agrees with the pressure $p_V$ obtained by the virial  MVE  method. Apparently, the formulas disagree, but if we
consider only the first order expansion in the parameter $\tau/\gamma$ of the previous formula the two methods
are in  agreement.
We start by rewriting 
\be
p_t=\frac{N T_s}{V}\int_V d\rr_1\dots\int_V d\rr_N P_N(\rr_1,\dots,\rr_N) -\frac{1}{3 V} \sum_i 
\int d\rr_1\dots\int d\rr_N P_N(\rr_1,\dots,\rr_N) \, \rr_i \cdot \frac{\partial \Ha_1}{\partial \rr_i} 
\ee
and notice that to first order in $\tau/\gamma$ 
 we have
\bea
&&P_N(\rr_1,\dots,\rr_N)  \sum_{\beta j}  \Bigl( \frac{\tau}{\gamma}  \frac{\partial \UU_{total} (\rr_1,\dots,\rr_N) }{\partial r_{\beta j}}  \frac{\partial^2  \UU_{total} }{\partial r_{\alpha i} \partial r_{\beta j}} -T_s \frac{\partial}{\partial r_{\alpha i}} \ln \det \Gamma\
  \Bigr) \approx \nonumber\\
 &&  - T_s  \frac{\tau}{\gamma}   \sum_{\beta j} \Bigl\{ [ \frac{\partial }{\partial r_{\beta j}} P_N(\rr_1,\dots,\rr_N) ]  \frac{\partial^2  \UU_{total} }{\partial r_{\alpha i} \partial r_{\beta j}} +P_N(\rr_1,\dots,\rr_N) \frac{\partial^3  \UU_{total} }{\partial r_{\alpha i} \partial r_{\beta j}^2} \Bigr\}
\eea
so that the presure can be rewritten as
\bea
&&p_t \approx \frac{N T_s}{V}\int d\rr_1\dots d\rr_N P_N -\frac{1}{3 V} \sum_i 
\int d\rr_1\dots\int_V d\rr_N P_N(\rr_1,\dots,\rr_N) \, \rr_i \cdot \frac{\partial \UU_{int} }{\partial \rr_i} \nonumber\\
&&+\frac{T_s}{3 V}   \frac{\tau}{\gamma} 
\int d\rr_1\dots d\rr_N  \sum_{\alpha i}   \sum_{\beta j}  \frac{\partial }{\partial r_{\beta j}} [P_N(\rr_1,\dots,\rr_N)  \frac{\partial^2  \UU_{total}}{\partial r_{\alpha i} \partial r_{\beta j}} ]  r_{\alpha i}
\eea
and after an integration by parts and discarding a boundary term we obtain
\be
p_t=-\frac{1}{3 V} \langle\sum_i \rr_i \cdot \ \frac{\partial \UU_{int}}{\partial \rr_i} \rangle 
+\frac{T_s}{3 V}   
\int d\rr_1\dots d\rr_N  \sum_{\alpha i}    P_N(\rr_1,\dots,\rr_N) [\delta_{\alpha\beta}- \frac{\tau}{\gamma}  \frac{\partial^2  \UU_{total}}{\partial r^2_{\alpha i} }]   
\ee
After recognising that the first term is the internal virial, and the last term is the first order approximation  to the trace of the
inverse matrix $\Gamma$, w find the result eq. \eqref{virialbaiesiold} of the main text.

 \section{Derivative of effective free energy with respect to the interfacial area}
 \label{arearescaling}
 
   In this case, we shall assume that the system is contained in a  box of volume $V= L_x\times  L_y \times L_z $, with $L_x=L_y=\sqrt A$
   and in order
 to properly take the derivative with respect to the area, while keeping the volume $V$ fixed, we non-dimensionalize each particle
 coordinate by the transformation:
$$(x_i,y_i,z_i)\to (\sqrt{A} X_i,\sqrt{A} Y_i,\frac{V}{A}Z_i)  $$
and assume  $0\leq  (X_i,Y_i,Z_i)\leq 1$ and rewrite the partition function as
  \be
   Z_N  =V^N \Pi_i^N \int_0^1\int_0^1 \int_0^1 dX_i d Y_i  dZ_i \dots  \exp \Bigl(- \frac{\Ha(\sqrt{A}X_1, \sqrt{A} Y_1,\frac{V}{A} Z_1,\dots,)}{T_s} \Bigr)
   \label{probabilitydistr}
    \ee
where $\Ha$ now contains only the interparticle potentials and not the external potential which is already accounted for by the
limits of integration.   
    Now we differentiate the partition function with respect to the area at constant volume and obtain:
  \be
  \frac{\partial  Z_N}{\partial A}=  - \frac{1}{ T_s}  V^N  \Pi_i^N \int_0^1\int_0^1 \int_0^1  d X_1 d Y_i d Z_i \dots  \exp \Bigl(- \frac{\Ha( \rr_1,\dots, \rr_N)}{T_s} \Bigr) \sum_i \frac{\partial  \rr_i}{\partial A} \cdot  \frac{\partial \Ha( \rr_1,\dots, \rr_N)}{\partial  \rr_i}  
\ee
Since the derivatives of the coordinates with respect to $A$ are:
$
\frac{\partial }{\partial A}(x_i,y_i,z_i)=
(\frac{1}{2  A} x_i ,\frac{1}{2 A}y_i ,-\frac{z_i}{A})$ we find:
  \be
  \frac{\partial  Z_N}{\partial A}=  - \frac{1}{ T_s}    \int_V d\rr_1\dots d\rr_N   \exp \Bigl(- \frac{\Ha( \rr_1,\dots, \rr_N)}{T_s} \Bigr) \sum_i \Bigl[\frac{x_i}{ 2A}  \frac{\partial \Ha( \rr_1,\dots, \rr_N)}{\partial  x_i} +\frac{y_i}{ 2A}  \frac{\partial \Ha( \rr_1,\dots, \rr_N)}{\partial  y_i}  -\frac{z_i}{ A}  \frac{\partial \Ha( \rr_1,\dots, \rr_N)}{\partial  z_i} \Bigr]  
\ee

 Going back to the original coordinates and recalling that the $x$ and $y$ direction are equivalent we obtain the formula:

\be
\gamma=-T_s\frac{\partial \ln Z_N}{\partial A}
= \frac{1}{ A}   \langle\sum_i  \Bigl[ x_i \frac{\partial }{\partial  x_i}  -z_i \frac{\partial }{\partial  z_i} \Bigr]   \Bigl (U_{int} 
+\frac{\tau}{2 \gamma}
    \sum_k^N (\frac{\partial \UU_{int}}{\partial\rr_k} )^2 
    - T_s  \ln  |\det ( I+ \frac{\tau}{\gamma}  \nabla \nabla \UU_{int} | \Bigr)
\rangle  
\ee

In order to prove the equivalence of the thermodynamic surface tension with its expression given in   section \ref{planar}, we use the same expansion up to linear order in $\tau/\gamma$ as in section \ref{volumerescaling} and obtain:
\be
\gamma\approx \frac{1}{ A}  \int_V d\rr_1\dots d\rr_N   \sum_i  \Bigl\{ \Bigl[ x_i \frac{\partial U_{int} }{\partial  x_i}  -z_i \frac{\partial U_{int} }{\partial  z_i} \Bigr] P_N  
- T_s \frac{\tau}{\gamma}
 \sum_{\beta j}   \frac{\partial }{\partial r_{\beta j}} [P_N \frac{\partial^2  \UU_{int}}{\partial x_ i \partial r_{\beta j}} ]  x_i-  \frac{\partial }{\partial r_{\beta j}} [P_N \frac{\partial^2  \UU_{int}}{\partial z_ i \partial r_{\beta j}} ]  z_i \Bigr\} .
\ee
After an integration by parts we get the result   eq. \eqref{surfacetension1} of the main text.
  

\end{document}